\documentclass [12pt]{article}
\usepackage{times}
\usepackage{epsfig}
\usepackage{amsmath}
\newcommand{\elp}{\mbox{$\mathrm{e}^+$}}
\newcommand{\elm}{\mbox{$\mathrm{e}^-$}}

\newcommand{\net}{\mbox{$\mathrm{\nu}$}}
\newcommand{\netb}{\mbox{$\mathrm{\overline{\nu}}$}}

\newcommand{\pip}{\mbox{$\mathrm{\pi^+}$}}
\newcommand{\pim}{\mbox{$\mathrm{\pi^-}$}}

\newcommand{\ppb}{\mbox{$\mathrm{p\overline{p}}$}}

\newcommand{\kl}{\mbox{$\mathrm{K_{L}}$}}
\newcommand{\ks}{\mbox{$\mathrm{K_{S}}$}}

\newcommand{\km}{\mbox{$\mathrm{K^{-}}$}}
\newcommand{\kp}{\mbox{$\mathrm{K^{+}}$}}

\renewcommand{\d}{\mbox{$\delta$}}

\newcommand{\e}{\mbox{$\epsilon$}}

\newcommand{\lS}{\mbox{$\lambda_{\mathrm{S}}$}}
\newcommand{\lL}{\mbox{$\lambda_{\mathrm{L}}$}}

\newcommand{\be}{\begin{equation}}
\newcommand{\ee}{\end{equation}}
\newcommand{\bea}{\begin{eqnarray}}
\newcommand{\eea}{\end{eqnarray}}

\def\a{\alpha}
\def\ba{\tilde{\alpha}}
\def\b{\beta}
\def\bb{\tilde{\beta}}

\begin{document}

\begin{flushright}
hep-ph/9903458 \\
CERN-TH/99-80  \\
LPTENS-99-12 \\
\end{flushright}

\vspace{8mm}

\begin{center}
{\bf
DIRECT 
\mbox{\boldmath{${\cal T}$}}-VIOLATION MEASUREMENTS AND }\\
\vspace*{0.15 cm} 
{ \bf
\mbox{\boldmath{${\cal T}$}}-ODD EFFECTS  IN
DECAY EXPERIMENTS
}
\end{center}

\vspace*{0.25cm}

\begin{center}
{ \bf
L. Alvarez-Gaum\'e $^{a}$,
~C. Kounnas $^{a,b}$, ~S. Lola $^{a}$ ~and~
P. Pavlopoulos $^{c}$
}

\vspace*{0.6 cm}
$^{a}$
CERN Theory Division, CH-1211 Geneva,
Switzerland \\
\vspace*{0.25 cm}
$^{b}$
Ecole Normale Sup\'erieure 
24 rue Lhommond, \\
F-75231, Paris Cedex 05, France \\
\vspace*{0.25 cm}
$^{c}$
Institut $~$f\"ur Physik, University of Basle CH-4056, and \\
CPLEAR Collaboration, CH-1211 Geneva
Switzerland

\end{center}

\vspace*{0.4 cm}

\begin{center}
{\bf ABSTRACT}
\end{center}

\noindent
{\small 
Motivated by the recent experimental announcements
for direct measurements of time-reversal
non-invariance in the neutral kaon system, we make a comparative discussion
of the CPLEAR and KTeV measurements.
The most suitable way 
to consistently incorporate the mixing, the time evolution 
and the decays of kaons, is to describe the neutral kaon system
as a system with a non-Hermitean Hamiltonian.
In this framework, the physical (decaying) incoming and outgoing states
are distinct and belong to dual spaces.
Moreover, since they are 
eigenstates of the full Hamiltonian, 
they never oscillate.
This is directly manifest in the orthogonality conditions
of the physical states, which 
entirely determine the evolution of the kaon system.
Along these lines we conclude:
CPLEAR studies $K^0$-$\bar{K}^0$ oscillations,
a process where initial and final states
can be reversed,  the CPLEAR asymmetry being an effect
directly related to the definition
of time-reversal.
Conclusively, CPLEAR provides 
a direct measurement of  ${\cal T}$-violation 
without any assumption either on unitarity or on $CPT$-invariance.
The KTeV experiment studies in particular the process
$K_L \rightarrow \pi^+ \pi^- e^+ e^-$, 
where they measure a ${\cal T}$-odd effect.
However, using unitarity together with estimates of the
final state interactions, it should be
possible to determine whether this
effect can be identified with a genuine
${\cal T}$-reversal violation
.}

\vspace*{0.4 cm}

\begin{center}
{\small {\it { Talk given by S. Lola at the XXXIV$^{th}$ Rencontres 
de Moriond on \\
Electroweak Interactions and Unified Theories, Les Arcs, 13-20 March 1999}}}
\end{center}

\thispagestyle{empty}

\setcounter{page}{0}
\vfill\eject

\newpage

\section{\large \sl Introduction}

Recently, the CPLEAR experiment at CERN, reported the
first direct observation of time-reversal violation
in the neutral kaon
 system~\cite{at_paper}. This observation is
made by comparing the probabilities
of a $\bar{K}^0$ state transforming into a
$K^0$ and vice-versa. 
Moreover, the KTeV experiment at Fermilab,
similarly reported evidence for
${\cal T}$-violation in the decay
$K_L \rightarrow \pi^+ \pi^- e^+ e^- $.~\cite{KTeV}
In the present note, we will discuss the
experimental asymmetries used by both collaborations
and interpret their measurements on 
${\cal CP}$, ${\cal T}$ and/or
${\cal CPT}$-violation.

The  discrete symmetry properties of the 
neutral kaon system
have been extensively studied 
in the literature \cite{rev}.
To analyse this issue, in a consistent way 
one needs to study a system with a non-hermitean Hamiltonian.
This is clear, because of the following:
Although the physical kaons at rest coincide with
the strong interaction (strangeness) eigenstates
$ |K^0> = |d \bar{s}>$ and  $|\bar{K}^0> = |\bar{d} s>$,
the latter are not the eigenstates of the the full
Hamiltonian. Since however weak interactions do not
conserve strangeness (but also allow $K^0$--$\bar{K}^0$ oscillations)
 the full Hamiltonian
eigenstates, denoted by
$|K_S>$ and $|K_L>$, are different from the 
strangeness eigenstates, and obey the relations
\bea
H~|K_S> & = & \lambda_S~|K_S>~, ~~~~~|K_S(t)> = e^{-i \lambda_S t} |K_S>~,
\nonumber \\
H~|K_L> & = & \lambda_L~|K_L>~, ~~~~~|K_L(t)> = e^{-i \lambda_L t} |K_L>~,
\eea
with $\lambda_L = m_L - i \Gamma_L / 2$
and $\lambda_S = m_S - i \Gamma_S /2 $,
where $m_{S,L}$ denotes the masses of the
physical kaons and 
$\Gamma_{S,L}$ their decay widths.
The complexity of the eigenvalues,
implies the non-hermiticity of the full
Hamiltonian of the neutral kaon system.

Non-hermiticity of $H$
implies that the physical incoming 
and outgoing states ($|K_{S,L}^{in}>$
and $|K_{S,L}^{out}> $ $\equiv $  $  {<K_{S,L}^{out}|}^{\dagger}
$ respectively), are not identical, 
but instead belong to two distinct (dual) spaces \cite{AKLP}.
In the Heisenberg representation (where the
states are time-independent), the
physical incoming and outgoing states coincide with
the left- and right-eigenstates of the full Hamiltonian:
\bea
H~|K_{S,L}^{in}>  ~=~  \lambda_{S,L}~|K_{S,L}^{in}>~,~~~~~
<K_{S,L}^{in}|~H^{\dagger} ~=~ <K_{S,L}^{in}|~\lambda^*_{S,L} ~,
\nonumber \\
H^{\dagger}|K_{S,L}^{out}> ~=~  \lambda_{S,L}^*~|K_{S,L}^{out}>~,~~~~~
{}~<K_{S,L}^{out}|~H  ~=~ <K_{S,L}^{out}|~\lambda_{S,L} ~,
\eea
where 
\bea
|K_{S,L}^{out}> ~~\equiv~~ {<K_{S,L}^{out}|}^{\dagger}  ~~\ne  ~~
|K_{S,L}^{in}>, ~~~~~
<K_{S,L}^{in}| ~~\equiv~~
{|K_{S,L}^{in}>}^{\dagger} ~~ \ne ~~
<K_{S,L}^{out}|~.
\eea
Notice that {\em only if }
$H=H^{\dagger}$, 
$\lambda_{S,L} ~=~ \lambda^*_{S,L}$ and 
$|K_{S,L}^{out}> ~=~ |K_{S,L}^{in}>$,
thus the incoming and outgoing states are identical.
In the generic case ($H \neq H^{\dagger}$),
the time  evolution  of the incoming and
outgoing
states $|\Psi_I^{in}(t_i)>$ and $|\Psi_I^{out}(t_f)>$  are
obtained
from $|\Psi_I^{in}>$ and $|\Psi_I^{out}>$, using the
evolution operators $e^{-iHt_i}$ and
$e^{-iH^{\dagger}t_f}$ respectively:
\bea
|K_{S,L}^{in}(t_i)> ~=~ e^{-iH\,t_i}~|K_{S,L}^{in}>, ~~~~~
|K_{S,L}^{out}{(t_f)}> ~=~ e^{-iH^{\dagger}\,t_f}~|K_{S,L}^{out}>~.
\eea
{}From the above equations, follows the evolution of the conjugate
states:
\bea
<K_{S,L}^{in}(t_i)| ~=~ <K_{S,L}^{in}|~ e^{iH^{\dagger}\,t_i}, ~~~~~
<K_{S,L}^{out}(t_f)| ~=~ <K_{S,L}^{out}|~e^{iH \,t_f}~. 
\eea
An important point to stress here, is
that the physical incoming and outgoing eigenstates
have to obey {\it at all times} the orthogonality conditions
\cite{AKLP}
\bea
<K_I^{out}(t_f)|K_J^{in}(t_i)>
=<K_I^{out}|e^{-iH\Delta t}|K_J^{in}>
=e^{-i\lambda_I~\Delta t}~\delta_{IJ}~,
\eea
and in particular for $ \Delta t = 0$
\bea
<K_L^{out}|K_S^{in}> = 0~, ~~~~~~~~ <K_S^{out}|K_L^{in}> = 0~,
\nonumber \\
<K_S^{out}|K_S^{in}> = 1~, ~~~~~~~~ <K_L^{out}|K_L^{in}> = 1 ~.
\label{ortho}
\eea
unlike what has been stated in a wide part of the
literature.
These conditions express the fact that the Hamiltonian
eigenstates cannot oscillate to each-other at any
time, and therefore an initial $|K_{S}^{in}>$ may not
be transformed to a final
$|K_{S}^{out}>$. Moreover, it follows that
the inner products among incoming
(outgoing) states {\em do not obey}
the usual  orthogonality conditions
\bea
<K_{I}^{in}|K_J^{in}> \ne \delta_{IJ}~~~~{\rm and}
{}~~~~ <K_I^{out}|K_J^{out}> \ne \delta_{IJ}~. 
\eea 
Finally, in the basis of the states $K_L$ and $K_S$,
$H$ can be expressed in terms of a diagonal  $2 \times  2$ matrix
\bea
H = |K_S^{in}> \lambda_S  < K_S^{out}| +
|K_L^{in}> \lambda_L <K_L^{out}| ~,
\label{HAMDIAG}
\eea
where the unity operator $\bf 1$ takes the form:
\bea
{\bf 1}= \sum_{I=S,L}~|K_{I}^{in}><K_{I}^{out}|~.
\eea

\section{\large \sl Study of discrete symmetries in the neutral kaon system}
\noindent

Having clarified our formalism, 
we may now proceed to study particle-antiparticle
mixing in the neutral kaon system.
A convenient representation to study
the action of ${\cal CP}$,
${\cal T}$ and ${\cal CPT}$, is the
$K^0, \bar{K}^0$ (particle-antiparticle) base.
In this representation
\begin{eqnarray}
CP \; |K_0^{in}> & = & |\bar{K}^{in}_0> ~, \nonumber \\
T \; |K_0^{in}> & = & <K_0^{out}|  ~, \nonumber \\
CPT \; |K_0^{in}> & = & <\bar{K}_0^{out}|~. 
\label{EQ1}
\end{eqnarray}

Without loss of generality,
we can express the physical incoming states in terms of
$|K_0^{in}>$ and  $|\bar{K}_0^{in}>$ as:
\begin{eqnarray}
|K_S^{in}> & = & \frac{1}{N} \left ( \; (1+\a) \; |K_0^{in}> + \;
            (1-\a) \; |\bar{K}_0^{in}>  \right ) ~, \nonumber \\
|K_L^{in}> & = & \frac{1}{N} \left ( \; (1+\b) \; |K_0^{in}> - \;
            (1-\b) \; |\bar{K}_0^{in}>  \right ) ~,
\label{EQ2}
\end{eqnarray}
where $\a$ and $\b$ are complex variables
associated with $CP$, $T$ and $CPT$-violation
(usually denoted by $\epsilon_S$ and 
$\epsilon_L$ respectively),
and $N$ a normalization factor. Then, the respective
equations for the outgoing states are not independent,
but are determined by the orthogonality conditions 
for the physical states \cite{AKLP}
\begin{eqnarray}
<K_S^{out}| & = & \frac{1}{\tilde{N}} \left ( \;
(1-\b) <K_0^{out}| + \; (1+\b) <\bar{K}_0^{out}|  \right ) ~, \nonumber \\
<K_L^{out}| & = & \frac{1}{\tilde{N}} \left ( \;
(1-\a) < {K}_0^{out}| - \; (1+\a)  <\bar{K}_0^{out}| \right ) ~.
\label{EQ3}
\end{eqnarray}
where the normalisation 
factor $N$ can always be chosen  equal to
$N=\sqrt{2 (1 - \alpha\beta)}$ \cite{AKLP}
Using the equations(\ref{HAMDIAG}, \ref{EQ2}, 
\ref{EQ3}) the Hamiltonian can be expressed 
in  the basis of
$K^0,\bar{K}^0$ as
\bea
H =  \frac{1}{2} \left(
\begin{array}{cc}
( \lL + \lS) -
\Delta \lambda \frac{ \a-\b}{1-\a\b} &
{}~~~~~~~~~~
\Delta \lambda   \;
\frac{1 + \a\b}{1-\a\b}
+ \Delta \lambda \frac{ \a+\b}{1-\a\b} \\
 & \\
\Delta \lambda  \;
\frac{1 + \a\b}{1-\a\b}
- \Delta \lambda \frac{\a+\b}{1-\a\b} &
{}~~~~~
( \lL + \lS) +
\Delta \lambda \frac{\a-\b}{1-\a\b}
\end{array}
\right )~, 
\label{eqH}
\eea
where $ \Delta \lambda = \lambda_L - \lambda_S$.

From eq.(\ref{eqH}), we can identify
the $T$-, $CP$- and $CPT$- violating parameters. Indeed:

$\bullet$ Under {\underline {$T$--transformations}},
 $$
<K_0^{out}|H |\bar{K}_0^{in}> \; \leftrightarrow \;
<\bar{K}_0^{out}|H|K_0^{in}> ~,
$$
thus, the off-diagonal elements of $H$
are interchanged.
This indicates that the parameter $\epsilon \equiv (\a+\b)/2$,
which is related to the difference of the
off-diagonal elements of $H$,
measures the magnitude of  the $T$-violation\footnote{
$~2/N^2 \approx 1$, in the linear approximation.}.
\begin{equation}
  \frac{2}{N^2}~\e = \frac {
<K_0^{out}|H|\bar{K}_0^{in}> - <\bar{K}_0^{out}|H|K_0^{in}>
}{2 \; \Delta \lambda} ~ .
\label{epsilon1}
\end{equation}

$\bullet$ Under {\underline {$CPT$--transformations}},
$$
<K_0^{out}|H |{ K}_0^{in}>
\; \leftrightarrow \;   <\bar{K}_0^{out}|H |\bar{K}_0^{in}> ~,
$$
and therefore, the parameter
$\delta \equiv (\a-\b)/2$, related to the
difference of the diagonal elements of $H$,
 measures the magnitude
of $CPT$-violation.
\bea
\frac{2}{N^2}~ \d = \frac{<\bar{K}_0^{out}|H|\bar{K}_0^{in}> -
<K_0^{out}|H|K_0^{in}>}
{2 \; \Delta \lambda}  ~.
\eea

$\bullet$ Under {\underline {$CP$--transformation}},
$$
<K_0^{out}|H |K_0^{in}> \; \leftrightarrow \;
<\bar{K}_0^{out}|H |{\bar
K}_0^{in}> ~,
$$
and simultaneously
$$
<K_0^{out}|H |\bar{K}_0^{in}>\; \leftrightarrow \;    <{\bar
K}_0^{out}|H|K_0^{in}> ~,
$$
thus, {\it both} the diagonal and the off-diagonal elements of
$H$
are interchanged.
Then, the parameters $\a=\e+\d$ and $\b=\e-\d$, are the ones
which measure the  magnitude of $CP$-violation in the decays of
$K_S$ and $K_L$  respectively.

\section{\large \sl 
CPLEAR direct measurement of time-reversibility}

Having identified the 
${\cal CP}$, ${\cal T}$ and  ${\cal CPT}$-violating
operations, one may construct asymmetries that measure
discrete symmetry-violations. For instance,
a time-reversal operation interchanges
 initial and final states,
with identical positions and opposite velocities:
\bea
T~[~<\bar{K}_0^{out} (t_f)|K_0^{in}(t_i)>~]~
=~<K_0^{out} (-t_i)|\bar{K}_0^{in}(-t_f)> ~.
\eea 
Assuming time-translation invariance
\bea
T~[~<\bar{K}_0^{out} (t_f)|K_0^{in}(t_i)>~]~
=~<K_0^{out} (t_f)|\bar{K}_0^{in}(t_i)>~.
\eea 
The time evolution from $t_i$ to $t_f$ implies
that 
\bea
<\bar{K}_0^{out} (t_f)|K_0^{in}(t_i)>
& =  &
\frac{1}{N^2} ~ (1-\a) (1-\b) ~
( e^{-i \lambda_S \Delta t} - e^{-i \lambda_L \Delta t} ) ~,
\eea 
\bea
<K_0^{out}(t_f)|\bar{K}_0^{in}(t_i)>
& = &
\frac{1}{N^2} ~ (1+\a) (1+\b) ~
( e^{-i \lambda_S \Delta t} - e^{-i \lambda_L \Delta t} ) ~.
\eea
Then, by definition,  the magnitude of 
${\cal T}$-violation is
directly related to 
the Kabir asymmetry \cite{Kabir}
\begin{eqnarray} 
 A_T 
&=&
\frac{
|<K_0^{out}(t_f)|\bar{K}_0^{in}(t_i)>|^2
-|<\bar{K}_0^{out} (t_f)|K_0^{in}(t_i)>|^2
} {
|<K_0^{out}(t_f)|\bar{K}_0^{in}(t_i)>|^2
+|<\bar{K}_0^{out} (t_f)|K_0^{in}(t_i)>|^2
}~, \nonumber \\
& & \nonumber \\
& = & \frac{
| (1+\a) (1+\b) |^2 - | (1-\a) (1-\b) |^2}
{| (1+\a) (1+\b) |^2 + | (1-\a) (1-\b) |^2} 
 ~\approx~ 4~Re[\epsilon] ~,
\end{eqnarray}
which is time-independent.
Any non-zero value for $A_T$ signals a direct
measurement of
$T$-violation without any assumption about $CPT$ invariance.
Here, we should note that
in linear order in $\epsilon$ and $\delta$, the
approximate equality
\bea
 <K_S^{in}|K_L^{in}>+<K_L^{in}|K_S^{in}>
{}~ \approx ~ 4 Re~[\epsilon] ~,
\label{Tviol}
\eea
holds. This follows directly 
from the non-orthogonality of
the adjoint states $<K_S^{in}|$ and $<K_L^{in}|$
that is manifest in
the equations 
\bea
<K_S^{in}|K_S^{in}>&=& \frac{1+|\alpha|^2}{|1-\alpha\beta|} ~,~~~~~~
<K_L^{in}|K_L^{in}> ~=~ \frac{1+|\beta|^2}{|1-\alpha\beta|} ~,
\nonumber\\
<K_S^{in}|K_L^{in}>&=& \frac{\alpha^*+\beta}{|1-\alpha\beta|}
{}~,~~~~~~
<K_L^{in}|K_S^{in}> ~=~ \frac{\alpha+\beta^*}{|1-\alpha\beta|} ~.
\eea
However, although the time-reversal asymmetry can
{\em in the linear approximation} be expressed
in terms of only incoming states, the conceptual
issue of reversing the time-arrow for
any ${\cal T}$-violation measurement is unambiguous.
For this reason, the CPLEAR
collaboration searched for
${\cal T}$-violation through $K^0$-$\bar{K}^0$ oscillations,
a process where initial and final states can be interchanged.

CPLEAR produces initial neutral kaons with defined strangeness from
proton-antiproton annihilations
at rest, via the reactions
\begin{equation} \ppb \longrightarrow
 \biggl \{\begin{array}{l}
           K^{-} \pip K^0  \\
           K^+ \pim \bar{K}^0 ~,
    \end{array}
\end{equation}
and tags the neutral kaon strangeness at the production time
by the charge of the accompanying charged kaon.
Since weak interactions do not conserve
strangeness,  the $K^0$ and $\bar{K}^0$ may subsequently transform
into each-other via oscillations with $\Delta S = 2$.
The final strangeness of the neutral kaon
is then tagged through the semi-leptonic
decays 
\bea
K^0 & \rightarrow  &  e^+ \pi^- \nu ~, ~~~
\bar{K}^0  \rightarrow    e^- \pi^+ \bar{\nu} ~,\nonumber \\
K^0 & \rightarrow  &  e^- \pi^+ \bar{\nu} ~, ~~~
\bar{K}^0  \rightarrow    e^+ \pi^- \nu ~.
\eea
Among them, the first two are characterized by
$\Delta S = \Delta Q$ while the other two
are characterized by
$\Delta S = -\Delta Q$ and would therefore indicate
either (i) explicit violations of the 
$\Delta S = \Delta Q$ rule, or 
(ii) oscillations between
$K^0$ and $\bar{K}^0$ that 
even if $\Delta S = \Delta Q$ holds,
would lead at a final state
similar to (i) (with the ``wrong-sign'' leptons).
The CPLEAR  experimental asymmetry  is given by
\bea
A_T^{exp}  =
\frac{\overline{R}_+ ~ (\Delta t) - R_{-} ~(\Delta t)}
{\overline{R}_+ ~(\Delta t) + R_{-} ~(\Delta t)} ~, \nonumber
\eea
 with 
\bea
             \overline{R}_+ {}~(\Delta t)
& = &
| <e^+ \pi^- \nu(t_f)~ |\bar{K}_0^{in}(t_i)> \nonumber \\
& + & 
 <e^+ \pi^- \nu(t_f)~ |K_0^{in}(t_f)>
<K_0^{out}(t_f)~ |\bar{K}_0^{in}(t_i)> |^2 ~,
\eea
\bea
 R_{-} ~(\Delta t) & = &
| <e^- \pi^+ \bar{\nu}(t_f)~ |{K}_0^{in}(t_i)> \nonumber \\
& + & <e^- \pi^+ \bar{\nu}(t_f)~ |\bar{K}_0^{in}(t_f)>
<\bar{K}_0^{out}(t_f)~ |K_0^{in}(t_i)> |^2 ~.
\eea 
where the first term in each sum stands for
(i) and the second for (ii) (thus containing
the kaon oscillations multiplied by the matrix element
for semileptonic decays through
$\Delta S = \Delta Q$.
The experimental asymmetry $A_T^{exp}$ therefore,
besides $\epsilon$, also contains the parameters
$ x_{-}$  and $y$, where $x_{-}$ 
measures $\Delta Q = -\Delta S$, while
$y$ stands for ${\cal CPT}$ violation in
the decays.
\bea
A_T^{exp}= 4 Re \; [\epsilon] ~  -  2 Re ~[x_{-}]  - 2 Re ~ [y]~.
\eea
In the CPLEAR experiment, with the proper experimental normalisations,
the measured asymptotic asymmetry is \cite{CPLEARlast}:
\bea
\tilde{A}_T^{exp}= 4 Re \; [\epsilon] ~  -  4 Re ~[x_{-}]  - 4 Re ~ [y]~.
\eea

The average value of $\tilde{A}_T^{exp}$  was found 
to be  $= (6.6 \pm 1.6) \times 10^{-3}$, which
 is to be compared to  the recent CPLEAR measurement of
$ (Re~[x_{-}] + Re~[y]) = (-2 \pm 3) \times 10^{-4}$,
indicating that the measured asymmetry 
is related to the violation of time-reversal invariance.
Conclusively, CPLEAR made a direct measurement
of time-reversal violation, as we had already
stated \cite{AKLP}.
Similar arguments have been presented \cite{JN}, using the density matrix 
formalism for the description of the kaon system.

An interesting question to ask at this stage,
is what information one could obtain from previous measurements
plus unitarity \cite{bell-stein}. Unitarity
implies the relations
\bea
<K_L^{in} |K_S^{in}> & = & \Sigma_{f}
<K_L^{in} |f^{in}> <f^{out} |K_S^{in}> ~,
\nonumber \\
<K_S^{in} |K_L^{in}> & = & \Sigma_{f}
<K_S^{in} |f^{in}> <f^{out}| K_L> ~,
\eea
where $f$ stands for {\it all}
possible decay channels.
Making the additional assumption
that the final decay modes satisfy the relation
\, $|f^{in}> = |f^{out}> \equiv <f^{out}|^\dagger$ \,
(which is equivalent to making use of
$CPT$-invariance of the final state interactions), it is
possible to calculate the sum
$ <K_L^{in}| K_S^{in}> + <K_S^{in} |K_L^{in}> $,
by {\it measuring only the branching ratios of kaon decays}.
This is what can be done in $K_L$, $K_S$ experiments, where only
the {\it incoming kaon states} are used.
(Note here, however, that in the next section we discuss a 
${\cal T}$-odd asymmetry that can be measured in a 
single decay channel).
In the linear approximation,
this sum is equal
to $4 ~Re~[\epsilon]$
(see eq. (\ref{Tviol})).
However, this is an {\it indirect}
determination of $T$-violation, and
would not have been possible if invisible decays were
present. This is to be contrasted with
the results of CPLEAR, which  
use only one out of the possible decaying channels,
and does not rely at all on unitarity and or the knowledge
of other decay channels than the one used
in the analysis \cite{AKLP}.

\section{ \large \sl ${\cal T}$-odd effects versus ${\cal T}$-reversal
violation}

The KTeV experiment looks at the rare decay
$K_L \rightarrow \pi^+ \pi^- e^+ e^-$
of which they have collected more than 2000 events.
In particular, they measure the asymmetry in 
the differential cross section, with respect
to the angle $\phi$ between the pion and electron
planes \cite{sehgal}. To give to the angle an unambiguous sign,
they define $\phi$ according to
\bea
\sin\phi \cos \phi = (\vec{n}_e + \vec{n}_\pi) \cdot \hat{z} ~,
\eea 
where 
$ \vec{n}_e (\vec{n}_\pi)$ is the unit vector
in the direction 
$ \vec{p}_{e^{-}} \times \vec{p}_{e^{+}} $
$ (\vec{p}_{\pi^{-}} \times \vec{p}_{\pi^{+}} $),
and $\hat{z}$ is the unit vector in the direction
of the sum of the two pion momenta \cite{sehgal}.
A ${\cal T}$-odd observable
is one that changes sign under the reversal of all
incoming and outgoing three-momenta and polarisations.
By construction, $\phi$ satisfies this property.
The operation of ${\cal T}$-reversal,
involves in addition to the operations mentioned,
a flip of the arrow of time (i.e. exchanging
initial and final states). The KTeV collaboration
observes an asymmetry of nearly $14 \%$ about $\phi = 0$,
thus identifying a ${\cal T}$-odd effect.

The important issue is to assess when such an effect can be
interpreted as a direct measurement of 
${\cal T}$-reversal violation, since 
nowhere have the initial and final states been interchanged
\cite{AdR}.
The key ingredient that effectively allows one to
invert the arrow of time in such a process,
is the hypothesis of the unitarity of the
$ S $-matrix: $S S^{\dagger} = 1$. The $S$-matrix
can be written in terms of the $T$-matrix
for a process $i \rightarrow f$, as
\bea
S_{if} = \delta_{if} + i T_{if}~,
\eea
where a $delta$-function for energy-momentum conservation
is included in $T_{if}$. Unitarity now implies:
\bea
T^*_{fi} = T_{if} - i A_{if}~,
\label{uni}
\eea
where $T_{fi}$ is the amplitude for a process $f \rightarrow i$
(i,e exchanging initial and final states), and $A_{if}$ is
the absorptive part of the $i \rightarrow f$ process:
\bea
A_{if} = \sum_{k} T_{ik} T_{fk}^* ~,
\eea
and the sum extends over all possible on-shell
intermediate states. Taking the absolute square of
(\ref{uni}):
\bea
|T_{fi}|^2 = |T_{if}|^2 + 2 Im (A_{if} T_{if}^*) + |A_{if}|^2 ~.
\label{squar}
\eea
If $\tilde{\imath}$,  $\tilde{f}$ denote the initial and
final states with three-momenta and polarisations reversed, 
${\cal T}$-reversal invariance would imply
\bea
|T_{fi}|^2 = |T_{\tilde{\imath} \tilde{f}}|^2 ~,
\eea
and from (\ref{squar}) we can construct
\bea
|T_{if}|^2 - |T_{\tilde{\imath}\tilde{f}}|^2 
& = & - 2 Im (A_{if} T_{if}^*) - |A_{if}|^2  \nonumber \\
& +  &  ( |T_{fi}|^2 - |T_{\tilde{\imath}\tilde{f}}|^2 ) ~.
\label{BASIC}
\eea
The left-hand side of (\ref{BASIC}) is precisely a
${\cal T}$-odd probability, for instance the one measured by
KTeV. However on the right-hand side we have two
contributions. The first contribution
arises from the terms in the first line corresponding
to final-state interactions (for instance
the exchange of a photon between the $\pi$'s
and $e$'s) which can affect the dependence on the angle
$\phi$ and generate a ${\cal T}$-odd effect through 
${\cal T}$-reversal conserving interactions.
The other contribution, the last line of (\ref{BASIC}),
is a genuine ${\cal T}$-reversal violating
contribution.
To identify a ${\cal T}$-odd effect with a violation
of ${\cal T}$-reversal,
it is thus necessary to estimate the effect of the
final state interactions for the process concerned
and to determine how big these contributions are
with respect to the measured ${\cal T}$-odd effect.
If these effects are small, then we can say that
using unitarity (and ${\cal CPT}$ invariance of the
final state interactions, which results in
$<\pi^+ \pi^- e^+ e^- |^{out} = 
(|\pi^+ \pi^- e^+ e^- >^{in})^\dagger $
), we are effectively interchanging the roles
of past and future and it is legitimate to
identify the ${\cal T}$-odd effect with a measurement
of ${\cal T}$-reversal violation.

\section{\large \sl Conclusions}

In the light of the recent data by
the CPLEAR and KTeV collaborations,
we discuss violations of discrete symmetries in the
neutral kaon system, with particular
emphasis to ${\cal T}$-reversal violation
versus ${\cal T}$-odd effects.
Since decaying kaons correspond mathematically
to a system with a non-hermitean Hamiltonian,
we use the dual space formalism, where
the physical (decaying) incoming and outgoing states
are distinct and dual of each-other.
This reflects the fact that the 
eigenstates of the full Hamiltonian 
may never oscillate to each-other and
have to be orthogonal at all times.
The orthogonality conditions
of the physical states, 
entirely determine the evolution of the kaon system.
In this framework, we study both the
asymmetries reported by CPLEAR and
KTeV and conclude the following:
CPLEAR, through $K^0$-$\bar{K}^0$ oscillations,
effectively reverses the arrow of time and thus
its measured asymmetry 
is directly related to the definition
of ${\cal T}$-reversal.
Having measured in the same experiment that additional
effects which enter in the experimental asymmetry
(arising by tagging the final kaon strangeness
by semileptonic decays, i.e.
violations of the $\Delta S = \Delta Q$ rule
and ${\cal CPT}$- invariance in the decays) are small,
it is concluded that CPLEAR indeed made the first
direct measurement of  ${\cal T}$-violation.
Since  the experiment
uses only one out of the possible decaying channels,
its results are also independent of any
unitarity assumption,
and the possible existence of invisible decay modes.

On the other hand, KTeV  studies the decay
$K_L \rightarrow \pi^+ \pi^- e^+ e^-$, which 
being an irreversible process
measures ${\cal T}$-odd effects.
These are not necessarily the same as
${\cal T}$-violating effects,
since they reverse momenta and polarisations but
not the time-arrow. It is straightforward to
demonstrate that ${\cal T}$-odd and ${\cal T}$-violating effects
are two different concepts.
Non-vanishing ${\cal T}$-odd effects
due to final state interactions,
may arise even if unitarity and ${\cal T}$-invariance hold.
However, since unitarity implies the inversion of the arrow of time,
a ${\cal T}$-odd effect could be interpreted as
time-reversal violation,  provided
${\cal CPT}$-invariance of the final states holds
and final state interactions are negligible.

\vspace*{0.3 cm}

\noindent
{\large \bf Acknowledgements: }
We would like to thank A. de Rujula, for very illuminating
discussions on ${\cal T}$-odd effects.
The work of C.K. is supported by the TMR contract
ERB-4061-PL-95-0789.


\end{document}